\documentstyle[emulateapj,apjfonts,psfig]{article}

\lefthead{J.~M.~Miller et al.}  \righthead{Bright ULXs in the Antennae}

\received{Date}
\revised{Date}
\accepted{Date}

\journalid{vol}{date}
\articleid{1}{4}
\paperid{id}

\cpright{AAS}{1999}
\ccc{x}

\begin{document}

\title{XMM-Newton Spectroscopy of Four Bright ULXs in the Antennae
  Galaxies\\ (NGC 4038/4039)}

\author{J.~M.~Miller\altaffilmark{1,2}, 
	A.~Zezas\altaffilmark{1},
	G.~Fabbiano\altaffilmark{1},     
     	F.~Schweizer\altaffilmark{3}
}

\altaffiltext{1}{Harvard-Smithsonian Center for Astrophysics, 60
        Garden Street, Cambridge, MA 02138, jmmiller@head-cfa.harvard.edu}
\altaffiltext{2}{National Science Foundation Astronomy and
        Astrophysics Fellow}
\altaffiltext{3}{Carnegie Observatories, 813 Santa Barbara St.,
Pasadena, CA 91101-1292}

\keywords{galaxies: individual (NGC 4038/4039) -- Black hole physics
-- X-rays:galaxies -- galaxies: stars}

\authoremail{jmmiller@cfa.harvard.edu}

\label{firstpage}

\begin{abstract}
We report the results of spectral fits to four bright ultra-luminous
X-ray sources (ULXs) in the Antennae galaxies (NGC 4038/4039) observed
for 41~ksec with {\it XMM-Newton}.  Although emission regions are not
resolved as well as in prior {\it Chandra} observations, at least four
ULXs (X-11, X-16, X-37, and X-44 in the Zezas \& Fabbiano scheme) are
sufficiently bright and well-separated with {\it XMM-Newton} that
reliable extractions and spectral analyses are possible.  We find that
the single component multi-color disk blackbody models cannot describe
any of the spectra.  Sources X-11 and X-16 are acceptably fit with
simple power-law models.  A thermal bremsstrahlung model provides a
better fit to the spectrum of X-44.  Including a disk blackbody
component to the spectrum of X-37 improves the fit and reveals an
apparently cool disk ($kT = 0.13 \pm 0.02$~ keV).  This would suggest
a parallel to cool disks recently found in other very luminous ULXs,
which may contain intermediate mass black holes; however, the complex
diffuse emission of the Antennae demands this finding be regarded
cautiously.

\end{abstract}

\section{Introduction}
ULXs are off-nuclear point-like X-ray sources in nearby normal
galaxies (Fabbiano \& Trinchieri 1987, Fabbiano 1989).  The X-ray flux
variability observed in many ULXs (see, e.g., La Parola et al.\ 2001,
Kubota et al.\ 2001) signifies that the class is likely
dominated by accreting compact objects.  The high luminosities
measured from ULXs by definition exceeds the Eddington luminosity for
a neutron star accreting isotropically; many exceed (sometimes by a
factor of 10 or more) the Eddington luminosity expected for a
$10~M_{\odot}$ black hole.
 
Theoretical models for ULXs suggest that these sources may be
stellar-mass sources with relativistically-beamed emission (Koerding,
Falcke, \& Markoff 2002), stellar-mass sources with anisotropic
emission (King et al.\ 2001), or intermediate-mass black holes
(IMBHs).  Observations with {\it Chandra} and {\it XMM-Newton} --- and
especially multi-wavelength observing schemes --- are beginning to
reveal that no single model for ULXs may describe the entire class.
Rather, the class may include examples of each possibility described
above (see recent reviews by King 2003, Fabbiano \& White 2003 and
Miller \& Colbert 2003).  NGC 5408 X-1 may be an example of a stellar
mass source with relativistically-beamed emission (Kaaret et al.\
2003); it is also possible that the spectrum can be described in part
with a cool disk, which would indicate an IMBH.  Many of the ULXs in
the Antennae (there are 9 sources with $L_{X} \geq 10^{39}$~erg/s for
a distance of 19~Mpc; Fabbiano, Zezas, \& Murray 2001) may be
stellar-mass sources with anisotropic emission, because their average
distances from the nearest star clusters suggest run-away binaries and
tend to exclude very massive systems, although capture of a companion
from a primordial IMBH cannot be excluded (Zezas \& Fabbiano 2002).
Finally, sources with high luminosities but low disk temperatures may
be IMBHs; NGC 1313 X-1 and X-2 (Miller et al.\ 2003a) may be examples.
 
At present, the anisotropic emission model (King et al.\ 2001) may be
the most compelling description for many ULXs.  It is motivated in
part by the unusually large number of ULXs resolved in the Antennae
with {\it Chandra}, that suggest a link with the actively forming
stellar population. This conclusion is supported by the association of
large numbers of ULXs with active star-forming galaxies (e.g. NGC
3256, Lira et al.  2001; NGC 4485/90 Roberts et al. 2002; NGC 1068,
Smith \& Wilson 2003).  However, the King et al.\ (2001) scenario may not
be able to describe sources at the highest end of the ULX luminosity
distribution (e.g., the sources with cool accretion disks mentioned
above; see also Strohmayer \& Mushotzky 2003).

In this paper we report the result of {\it XMM-Newton} observations of
the Antennae, which provide for the first time high-quality
medium-resolution X-ray spectra of four bright ULXs in this merger
system.  These data can be used to constrain the emission model, as it
was done in the case of the discovery of low-temperature components in
the NGC~1313 ULXs (Miller et al.\ 2003).  The spatial resolution of
{\it XMM-Newton} is approximately ten times more coarse than that of
{\it Chandra}, so that robust spectra can only be obtained from the
brightest sources in targets with crowded fields like the Antennae.
The procedure used to reduce our data and extract spectra is described
in Section 2.  In Section 3, we present the results of fitting simple
spectral models to four bright ULXs.  Finally, in Section 4, we
discuss our results and their impact on our understanding of ULXs
within the context of models for their nature and anomalous
luminosity.

\section{Data Reduction and Analysis}
The Antennae (NGC 4038/4039) were observed with {\it XMM-Newton} for
40.9 ksec on 8 January 2002, starting at 22:03:37 (UT).  We used only
the EPIC data for this analysis.  The EPIC MOS-1 and MOS-2 cameras
were operated in ``PrimeFullWindow'' mode and the EPIC-pn camera was
operated in ``PrimeFullWindowExtended'' mode.  The  ``medium''
optical blocking filter was used for for all three cameras..

The {\it XMM-Newton} reduction and analysis
suite SAS version 5.3.3 was used to filter the standard pipeline event
lists, detect sources within the field, and make spectra and
responses.  After filtering against soft proton flares, the net ``good
time'' for each camera was 22.6 ksec, 25.6 ksec, and 25.7 ksec for the
pn, MOS-1, and MOS-2 cameras, respectively.

The ULX population in the Antennae is concentrated near the center of
the interacting galaxies, which is a region with a dense source
population.  By relying upon the {\it Chandra} source detections, we
determined that many sources are blended or too close to other bright
sources to reliably extract a spectrum.  Fortunately, at least four
ULXs are well-separated from others, and/or bright enough that counts
within a reasonable extraction region should be strongly dominated by
the bright source.  These sources are X-11, X-16, X-37, and X-44 (as
numbered and identified by Zezas et al.\ 2002a) and are shown in
Figure 1.

We used circular regions centered on the known {\it Chandra} positions
to extract source counts for these ULXs.  The recipes described in the
MPE ``cookbook'' were used to filter the event lists.  These
parameters were chosen as follows: we set ``FLAG=0'' to reject events
from bad pixels and events too close to the CCD chip edges; patterns
0-4 were selected for the pn camera, and patterns 0-12 for the MOS-1
and MOS-2 cameras; finally, the pn spectral channels were grouped by a
factor of 5 and the MOS-1 and MOS-2 channels by a factor of 15.

For X-11, X-16, and X-44, source counts were extracted within a 12''
radius; due to the proximity of a nearby source a 9'' extraction
radius was used for X-37.  Flux from source X-13 --- a transient soft
source (Fabbiano et al.\ 2003b) --- might contaminate the flux
extracted from X-11.  However, no soft emission component is required
to describe the spectra of X-11 (see Section 3), so it is likely that
X-13 was in a low flux state.  Background spectra were not extracted
for two reasons.  First, for {\it XMM-Newton}, approximately 40\% of
the on-axis point source encircled energy lies between radii of 15 and
50 arcseconds.  Extracting a background spectrum in an annulus for
subtraction from the source spectrum would primarily act to subtract
source counts.  Second, the central region of the Antennae galaxies is
a crowded field; adjacent regions may include emission from point
sources that are weaker than the brightest ULXs and prevent
measurement of a true (diffuse) background.

Observations with {\it Chandra} suggest that the diffuse background
near to the brightest ULXs should be less than 1\% of the source
counts, however, and so should not significantly affect fits to the
source spectra.  Moreover, sources resolved with {\it Chandra} that
are nearby to the four brightest ULXs considered in this work were
previously found to generally be one to two orders of magnitude
fainter than the ULXs (Zezas et al.\ 2002a).  Given the fall-off in
encircled energy noted above, flux from nearby point sources is likely
to be less than 5\% of the flux in the ULX extraction regions.  See
Section 3, however, for a brief discussion of the observed local
background, which is more complex than these simple estimates would
indcate.  Response files were generated for each spectrum from each
camera using the SAS tools ``rmfgen'' and ``arfgen''.  Prior to
fitting, the spectra were grouped to require at least 10 counts per
bin to ensure the validity of $\chi^{2}$ statistics.  Grouping to
require more counts per bin did not significantly change the fit
results.

The ULX light curves (in the 0.3--10.0 keV band, with 1.0 ksec bins)
are well-described by a constant flux level.  We therefore considered
only the time-averaged spectra from each ULX.  Model spectra were fit
to the data using XSPEC version 11.2 (Arnaud 1996).  The pn, MOS-1,
and MOS-2 spectra of each source were fit jointly, allowing an overall
normalizing constant to account for small differences in the
broad-band flux calibration of the cameras.  Values obtained for the
constant indicate that at the time of observation, the pn and MOS-1
flux calibrations differ by less than 5\%.  The MOS-2 flux calibration
differed by less than 5\% from the MOS-1 camera, but may have differed
from the pn at the 10\% level.  Systematic errors were not added to
the spectra prior to fitting.  Models were fit to the EPIC spectra in
the 0.3--10.0 keV band.  Errors quoted in this work are at the 90\%
confidence level for one interesting paramter.  Using the SAS tool
``epatplot'' and the HEASARC tool ``PIMMS'' we found that the effects
of photon pile-up were negligible in this observation.

\section{Results}
An examination of the individual spectra did not reveal compelling
evidence for emission or absorption lines; where a possible feature is
seen in one spectrum (pn, MOS-1, or MOS-2), in each case there is no
feature evident in the other spectra.  This strongly suggests that any
narrow deviations from a continuum spectrum are statistical
fluctuations.  We therefore restricted our modeling to continuum
components.  For each model, the ``phabs'' model in XSPEC was used to
describe the equivalent neutral hydrogen column density along our line
of sight.  We fixed the lower bound of the equivalent neutral hydrogen
column density to the Galactic value ($N_{H} = 4.0 \times 10^{20}~{\rm
cm}^{-2}$, Dickey \& Lockman 1990).  The results of fits with a number
of simple models are listed in Table 1, and the results are described
in detail below.

Prior fits to a number of ULX spectra (not sources within the
Antennae) obtained with {\it ASCA} with the multicolor disk (MCD)
blackbody model (Mitsuda et al.\ 1984) suggested that hot accretion
disks may be a fundamental accretion flow structure in ULXs
(Makishima et al.\ 2000).  Fits to X-11, X-16, X-37, and X-44 yielded
high disk color temperatures ($1.0 {\rm keV} < kT < 1.7 {\rm keV}$),
similar to prior {\it ASCA} results.  However, this model does not
yield an acceptable fit to any of the spectra.  Other single-component
models and two-component models are strongly preferred statistically
(see below).  This may suggest that X-11, X-16, X-37, and X-44 may be
different from ULXs such as Dwingeloo X-1 (see Makishima et al.\
2000).  It is possible, however, that the apparent differences are
partially due to the better sensitivity of {\it XMM-Newton}.

Significantly better fits are obtained when the spectra are fit with
either a power-law or thermal Bremsstrahlung model.  In systems
accreting at high rates, most of the flux phenomenologically
described as a power-law likely arises from the Compton-upscattering
of relatively cool disk photons in an optically-thin corona (see,
e.g., Titarchuk 1994).  A thermal Bremsstrahlung spectrum is
consistent with expectations for an optically-thick, {\it
geometrically} thick disk which may have a significant advective
component (see, e.g., Ebisuzaki et al. 2001, Begelman 2002).
Statistically acceptable fits to the spectra of X-11 and X-16 are
obtained with either continuum model (see Table 1).  The spectra of
X-37 are not acceptably described with either model.   A bremsstrahlung
model is a better description of the spectra from X-44 than a
power-law model.  The fits to X-11, X-16, X-37, and X-44 with a
power-law model and associated data/model ratios are shown in Figures
2--5.  The best-fit models for the spectra from each ULX are shown in
Figure 6.

The addition of a simple blackbody component to a power-law yielded
cool temperatures and marginal statistical improvements to the fits.
However, disk temperatures certainly fall off with radius (see, e.g.,
Frank, King, \& Raine 2002), and are more properly modeled as a series
of blackbody annuli as approximated by the MCD model.  A single
blackbody component might reasonably approximate a neutron star
surface; however, the luminosities inferred in these sources are two
orders of magnitude above the Eddington limit for a neutron star,
which is the basis for the the often-made assumption that ULXs harbor
accreting black holes.  

Similarly, the addition of a low-temperature diffuse thermal emission
component (e.g., a MEKAL plasma) to a power-law gives marginal
statistical improvement when fitting the ULX spectra.  However, this
model is likely not justified.  Observations of black hole binaries in
the Milky Way and Magellanic Clouds do not reveal evidence of
optically-thin thermal emission local to the systems through plasma
emission line spectra.  It is particularly interesting to note that
such spectra have not been revealed even in systems with high mass (O
or B type) companions, given the emerging links between ULXs and
star-formation (for reviews, see Fabbiano \& White 2003 and Miller \&
Colbert 2003).  Observations of Cygnus X-1 (O9.7 Iab companion, Gies
\& Bolton 1982) with the {\it Chandra} High Energy Transmission
Grating Spectrometer have revealed a rich {\it absorption} spectrum in
all states (Schulz et al. 2002, Marshall et al.\ 2002, Miller et al.\
2003b, Feng et al. 2003).  LMC X-1 (O7 III companion, Cowley et al.\
1995) is often observed at luminosities exceeding $10^{39}~{\rm
erg}~{\rm s}^{-1}$ (Wilms et al.\ 2001), but even {\it Chandra}/HETGS
spectra have failed to reveal emission lines from an optically-thin
plasma (Cui et al.\ 2002).  Moreover, the diffuse emission in the
Antennae which can be described in terms of optically-thin thermal
models (Fabbiano et al.\ 2003a) is expected to be two orders of
magnitude less than the flux from a ULX.

Finally, we considered a model consisting of MCD and power-law
components.  This simple model is often applied to stellar-mass black
hole binaries in the Milky Way and Magellanic Clouds.  Moreover, as
noted above, it has a stronger physical motivation than simple
blackbody plus power-law or optically-thin thermal emission plus
power-law models.  The MCD plus power-law model permits an interesting
comparison to recent ULX spectral results in which MCD color
temperatures 5--10 times lower than commonly-measured in stellar-mass
black holes may imply IMBHs since $T \sim M_{BH}^{-1/4}$ in the MCD
model (Miller et al. 2003; see also Kaaret et al. 2003).  While this
model is not required to describe the spectra of X-11 or X-16, it
provides a marginally better fit, and represents a statistically
better fit to X-37 at the $3.5\sigma$ level of confidence based on an
F-test.  However, this two component model provides no improvement for
X-44.  In each case, the color temperature obtained is consistent with
a cool disk (the range of temperatures obtained is $0.11 {\rm keV} <
kT < 0.21 {\rm keV}$).  The power-law component represents a
significant fraction of the 0.3--10.0~keV flux in this model.  This
may suggest that the gas in any optically-thin accretion flow geometry
is very hot; if so, it is unlikely that the potentially cool thermal
emission we have characterized with a disk model is actually due to
local illumination of diffuse gas.  This two-component model is the
best description of the spectra of X-37; for this source we measure
$kT = 0.13 \pm 0.02$~keV.

If color temperatures of $kT \sim 0.5-1.0$~keV may be taken as typical
of stellar-mass black holes near to or above $L_{X}/L_{Edd.} \sim
0.1$, then color temperatures in the $kT \sim 0.11-0.21$~keV range may
indicate black holes with masses in the range of $30~M_{\odot} \leq
M_{BH} \leq 6800~M_{\odot}$ in these ULXs.  The two-component MCD plus
power-law model provides the best fit to X-37; for this source, black
hole masses in the range $120 M_{\odot} \leq M_{BH} \leq 7000
M_{\odot}$ (90\% confidence, including uncertainties in $kT_{X-37}$
and stellar-mass black hole disk color temperatures in the range $0.5
{\rm keV} \leq kT \leq 1.0 {\rm keV}$) may be suggested.  It is worth
noting, however, that the model used represents only a $3.5\sigma$
improvement to the spectra of X-37, far less than the $8\sigma$
improvement over single-component fits to NGC 1313 X-1 ($kT_{NGC 1313
X-1} = 0.15^{+0.02}_{-0.04}$ keV; Miller et al. 2003).

Given the impact that cool soft components can have on models for the
nature of a given ULX, it is worth carefully re-examining the question
of local background contamination.  The Antennae are known to have a
complex, spatially variable diffuse emission component (Fabbiano et
al.\ 2004).  To investigate this more fully and to test our
assumptions regarding background contamination (see Section 2), we
extracted a number of local background regions inside of, along, and
outside of the core ring of emission seen in Figure 1.  Outside of
this ring and in some locations within the ring, the total background
(in counts per area) is only 3--8\% of the flux in our source
extraction regions.  In other locations -- particularly at points
along the inner edge of the ring -- the background can be as high as
20--30\% of flux in our source extraction regions.  The intensity and
spectrum of the local background varies considerably, consistent with
Fabbiano et al.\ (2004).  In view of the complex diffuse background,
the possibility of a cool disk component in X-37 must be regarded more
cautiously than in cases like NGC 1313 X-1 and X-2.

\section{Discussion}
We have analyzed the {\it XMM-Newton}/EPIC spectra of four bright ULX
sources (X-11, X-16, X-37, and X-44; as numbered by Zezas et al.\
2002a).  The spectra of two sources (X-16, X-11) can be well described
by a single power-law ($\Gamma_{X-11}=1.48$, $\Gamma_{X-16}=1.9$). One
of the other sources (X-44) is best described by a thermal
bremsstrahlung model ($kT=3.7$~keV) while the fourth source (X-37)
may require two spectral components: a power-law ($\Gamma_{X-37}=2.0$)
and a MCD model ($kT_{X-37}=0.13$~keV).  Here we comment on the
implications of these results for the nature of these sources and the
general ULX population.

The coadded spectrum of the ULXs detected in the first {\it Chandra}
observations of the Antennae was well represented by a composite
MCD/power-law model with a temperature of 1.13~keV (Zezas et al.\
2002b), while similar fits of the {\it XMM-Newton} spectra give
temperatures in the $0.1-0.2$~keV range.  This difference is most
probably due to the fact that the {\it Chandra} results are for the
coadded spectra of 18 sources, while here we study the detailed
spectra of individual sources.  In fact a comparison with the
individual {\it Chandra} spectral fits for these sources (Zezas et
al.\ 2002b) shows that the photon indices from the {\it Chandra}
power-law fits are slightly flatter, but consistent within the
$2~\sigma$ level, with those derived from the {\it XMM-Newton} fits.
Similarly, the values of $N_{H}$ from the {\it Chandra} fits are
slightly higher than those derived from the {\it XMM-Newton} fits, but
again consistent within the $2\sigma$ level.  The discrepancy in
$N_{H}$ may be due to the fact that the degradation of the {\it
Chandra}/ACIS quantum efficiency was not taken into account by Zezas
et al.\ (2002b).  We note that no flux variability above the $3\sigma$
level has been detected between the two observations.

The thermal bremsstrahlung model that best describes the spectrum of
X-44 is broadly consistent with expectations for a ``slim'' disk
around a stellar-mass black hole (Watarai, Mizuno, \& Mineshige 2001),
or perhaps a thin disk radiating above the Eddington limit due to
photon bubble instabilities (Begelman 2002).  Additional sensitive
observations of X-44 may be able to confirm or reject these disk
models by determining if the spectrum we have measured is typical.

The spectra of the other ULXs are even more difficult to parse.  The
power-law indices measured from X-11, X-16, and X-37 are well within
the range commonly observed in Galactic stellar-mass black holes (for
reviews, see, Tanaka \& Lewin 1995 and McClintock \& Remillard 2003).
If these sources were ``microblazars'' (e.g., Kording, Falcke, \&
Markoff 2002), in which a jet axis coinciding with the line of sight
might boost the flux through beaming, the same beaming should act to
create a significantly harder power-law spectrum.  In addition,
beaming should tend to create strong short-timescale flux variations,
which are also not seen in these sources.  While very sensitive
optical and radio limits are needed to more definitively rule-out a
microblazar interpretation (and, this is complicated greatly by the
distance of the Antennae galaxies), the X-ray properties of these
sources are certainly inconsistent with a microblazar scenario.

Another possibility is that X-11, X-16, and X-37 are stellar-mass
black holes in the ``very high'' state (see the reviews mentioned
previously; see also Kubota, Done, \& Makishima 2002).  Certainly, if
ULXs behave like Galactic stellar-mass black holes, it is reasonable
to expect that this state holds at the luminosities observed.  Whereas
Cygnus X-1 gets softer when it brightens in the soft X-ray band (below
10~keV), X-11 and X-16 have been observed to get harder when they get
brighter (Fabbiano et al.\ 2003b); this is sometimes referred to as
microquasar-like variability since the very high state is spectrally
harder than the high/soft state.  However, at high accretion rates a
hot disk component ($kT \simeq 1$~keV, or higher) is always important,
even in cases where the power-law component appears to dominate (e.g.,
Zycki, Done, \& Smith 1999).  The spectra of X-11, X-16, and X-37
cannot be described by fits consisting only of hot MCD components.
Moreover, as noted previously, when MCD plus power-law models are fit
to the spectra, cool disks temperatures are measured.  Identifying
X-11, X-16, and X-37 as stellar-mass black holes in a very high state
similar to that defined in Galactic systems is therefore very problematic.

The $kT = 0.13\pm 0.02$~keV disk component in the spectrum of X-37 ---
if due to a standard disk around an IMBH --- implies a mass in the
$120 - 7000~M_{\odot}$ range.  Similar temperatures have been measured
in the spectra of NGC 1313 X-1 and NGC 1313 X-2 (Miller et al.\
2003a), and NGC 5408 X-1 (Kaaret et al.\ 2003).  As noted in Section
3, however, it is possible that the complex, spatially variable
diffuse emission in the Antennae has contaminated the spectrum of
X-37; part of the soft component may be background flux.  The spatial
resolution afforded by {\it XMM-Newton} makes it unlikely that future
observations with {\it XMM-Newton} will determine the nature of the
apparent soft component in X-37 with greater certainty.  Analysis of
new, long {\it Chandra} observations, or future {\it Chandra}
observations, may be able to make such a determination.

A comparison of the {\it Chandra} position of X-37 and the positions
of its nearby star-clusters shows that that there are no star-clusters
down to $m_{V}=23~mag$ within a radius of 1.7'' (Zezas et al.\ 2002b),
which translates to a physical distance of 156~pc ($d=19.0$~Mpc).
Following Zezas et al.\ (2002b), if this separation is due to a
supernova kick to the system, and assuming that {\it a}) conservation
of momentum, {\it b}) the same mechanism imparting kicks to neutron
star systems is active here, and {\it c}) the X-ray binary becomes
active in the end of the donor's main sequence lifetime, we can set an
upper limit to its mass of $\sim6~M_{\odot}$ if the compact object has
a mass of $120~M_{\odot}$; this corresponds to a star of spectral type
B5 or later.  However, the large mass ratio between the putative IMBH and the
donor makes this type of system prone to transient behavior.
Therefore, if this object is an IMBH, then there may be a 
large population of such systems throughout the Antennae (relatively
few may be observed at any given moment, depending upon their duty
cycles).

King and Pounds (2003) have proposed a model wherein the cool thermal
emission revealed in some ULXs may be due to an optically-thick,
quasi-spherical outflow originating approximately 100 Schwarzschild
radii from a stellar-mass black hole accreting at a rate close to the
Eddington limit.  As noted previously, when stellar-mass black holes
are near peak luminosity, they are found in the very high state.  A
number of hallmark features of the very high state clearly reveal the
inner disk, and are plainly inconsistent with an optically-thick
outflow obscuring the innermost accretion environment (for a recent
review, see McClintock \& Remillard 2003).  First, a hot thermal
accretion disk component ($kT \simeq 1$~keV, or above) is always
observed in the very high state.  Second, high frequency (${\rm few}
\times 100$~Hz) quasi-periodic oscillations (QPOs) --- which are
connected to the Keplerian orbital frequency in the inner disk, to
resonances in the inner disk in a Kerr spacetime, or both --- are
preferentially found in the very high state (some are also found in
the intermediate state, which is likely a lower-flux version of the
very high state).  Third, broad, relativistic Fe~K$\alpha$ emission
lines due to hard X-ray emission irradiating the inner disk are
preferentially revealed in the very high state.  The model suggested
by King \& Pounds (2003) also requires any hard X-ray emission to
originate in shocks outside of the optically-thick outflow.  This is
again inconsistent with observations.  High frequency QPOs (and, lower
frequency QPOs) are a higher fraction of the rms noise at higher
energy, suggesting a fundamental connection between inner disk
frequencies and the hard X-ray emitting region.  Relativistic
Fe~K$\alpha$ emission lines also require the source of hard X-ray
emission to be very central in the accretion flow.  Fourth, the photon
index of the power-law components often observed in bright ULXs is
harder than is generally found in the very high state of Galactic
black holes (where $\Gamma \simeq 2.4$ is common, McClintock \&
Remillard 2003, Roberts et al.\ 2004).  Finally, high inclination
``dipping'' X-ray binaries clearly require that the hard component in
these systems be fairly compact and central.  Clearly, if the King \&
Pounds (2003) model is to hold, the source state required cannot be
like those already observed in Galactic black holes in their highest
flux phases.

In summary, the spectra we have obtained do not allow for definitive
conclusions regarding the nature of the four brightest ULXs in the
Antennae galaxies.  However, the spectra are of sufficient quality to
imply that the ULXs in the Antennae may not all have a common nature.
These results provide further evidence that models for the ULX
phenomenon based on a single source population may not be sufficient
to describe the entire class.  

J.M.M. acknowledges support from the NSF through its Astronomy and
Astrophysics Fellowship Program.  We thank the anonymous referee for
helpful suggestions.  This work is based on observations obtained with
{\it XMM-Newton}, an ESA mission with instruments and contributions
directly funded by ESA Member States and the US (NASA).  This work
made use of the High Energy Astrophysics Archive Research Center
(HEASARC), operated for NASA by GSFC.

\clearpage

\begin{table}[h]
\caption{Spectral Fit Parameters}
\begin{footnotesize}
\begin{center}
\begin{tabular}{llllll}
\multicolumn{2}{l}{Model/Parameter} & X-11 & X-16 & X-37 & X-44\\
\tableline

\multicolumn{2}{l}{power-law} & ~ & ~ & ~ & ~ \\
\multicolumn{2}{l}{$N_{H}~(10^{21}~{cm}^{-2})$} & $1.0 \pm 0.2$ & $0.4 \pm 0.2$ & $2.0 \pm 0.3$ & $1.3 \pm 0.3$ \\
\multicolumn{2}{l}{$\Gamma$} & $1.9 \pm 0.1$ & $1.48 \pm 0.07$ & $2.06 \pm 0.09$ & $2.2 \pm 0.2$ \\
\multicolumn{2}{l}{Norm. ($10^{-5}$)} & $5.5 \pm 0.7$ & $3.6 \pm 0.3$ & $5.3^{+0.9}_{-0.7}$ & $5 \pm 1$ \\ 
\multicolumn{2}{l}{$F_{0.3-10}~(10^{-13}~erg~cm^{-2}~s^{-1})^{a}$} &
$3.3 \pm 0.4$ & $2.9 \pm 0.2$ & $2.5^{+0.4}_{-0.3}$ & $2.3 \pm 0.5$ \\
\multicolumn{2}{l}{$\chi^{2}/dof$} & 162.6/160 & 140.0/164 & 163.0/129 & 162.6/139 \\
\tableline

\multicolumn{2}{l}{MCD} & ~ & ~ & ~ & ~ \\
\multicolumn{2}{l}{$N_{H}~(10^{21}~{cm}^{-2})$} & $0.4 \pm 0.1$ & $0.40 \pm 0.02$ & $0.4 \pm 0.1$ & $0.40 \pm 0.03$ \\
\multicolumn{2}{l}{$kT$~(keV)} & $1.1 \pm 0.1$ & $1.5^{+0.2}_{-0.1}$ & $1.2 \pm 0.1$ & $1.0 \pm 0.1$ \\
\multicolumn{2}{l}{Norm. ($10^{-3}$)} & $7.0 \pm 1.2$ & $2.2 \pm 0.7$ & $4.5^{+1.7}_{-1.2}$ & $9^{+2}_{-3}$ \\
\multicolumn{2}{l}{$\chi^{2}/dof$} & 245.8/160 & 258.8/164 & 220.2/129 & 194.8/139\\
\tableline

\multicolumn{2}{l}{Bremsstrahlung} & ~ & ~ & ~ & ~ \\
\multicolumn{2}{l}{$N_{H}~(10^{21}~{cm}^{-2})$} & $0.4 \pm 0.1$ & $<0.4$ & $1.1 \pm 0.2$ & $0.5^{+0.2}_{-0.1}$ \\
\multicolumn{2}{l}{$kT$~(keV)} & $5.1 \pm 0.8$ & $12^{+6}_{-2}$ & $4.5 \pm 0.8$ & $3.7 \pm 0.5$ \\
\multicolumn{2}{l}{Norm. ($10^{-5}$)} & $6.0 \pm 0.6$ & $5.2 \pm 0.4$ & $5.1 \pm 0.6$ & $4.9^{+0.7}_{-0.3}$ \\ 
\multicolumn{2}{l}{$\chi^{2}/dof$} & 164.1/160 & 152.1/164 & 178.7/129 & 149.4/139 \\
\tableline

\multicolumn{2}{l}{MCD $+$ power-law} & ~ & ~ \\
\multicolumn{2}{l}{$N_{H}~(10^{21}~{cm}^{-2})$} & $3.0^{+0.8}_{-1.8}$ & $1.5 \pm 1.0$ & $5.6 \pm 1.5$ & $1.4^{+2.0}_{-0.4}$ \\
\multicolumn{2}{l}{$kT$~(keV)} & $0.15 \pm 0.02$ & $0.19 \pm 0.05$ & $0.13 \pm 0.02$ & $0.15^{+0.02}_{-0.15}$ \\
\multicolumn{2}{l}{Norm.} & $50^{+10}_{-48}$ & $5^{+1}_{-4}$ & $250 \pm 150$ & $1^{+20}_{-1}$ \\
\multicolumn{2}{l}{$\Gamma$} & $1.9 \pm 0.2$ & $1.4 \pm 0.2$ & $2.0 \pm 0.2$ & $2.2^{+0.1}_{-0.4}$ \\
\multicolumn{2}{l}{Norm. ($10^{-5}$)} & $5.5^{+1.7}_{-1.0}$ & $3.2^{+0.9}_{-0.7}$ & $5.4 \pm 1.5$ & $5.0 \pm 1.0$\\
\multicolumn{2}{l}{$\chi^{2}/dof$} & 153.2/158 & 131.5/162 & 143.8/127 & 162.2/137 \\
\multicolumn{2}{l}{$F~(10^{-13}~erg~cm^{-2}~s^{-1})^{a}$} &
$2.5^{+0.7}_{-1.3}$ & $2.9^{+0.7}_{-1.8}$ & $1.9^{+0.9}_{-0.7}$ & $1.8^{+2.3}_{-0.4}$ \\
\multicolumn{2}{l}{$F~(10^{-13}~erg~cm^{-2}~s^{-1})^{b}$} &
$4.9^{+1.3}_{-2.5}$ & $3.8^{+0.9}_{-2.3}$ & $7.9^{+3.9}_{-2.8}$ & $2.3^{+3.0}_{-0.5}$ \\
\multicolumn{2}{l}{$F_{power-law}/F_{total}$} & 0.61 & 0.33 & 0.33 & 0.77 \\
\multicolumn{2}{l}{$L_{0.3-10}~(10^{40}~{\rm erg/s})^{c}$} & $2.1^{+0.7}_{-1.1}$ & $1.6^{+0.4}_{-1.0}$ & $3.4^{+1.7}_{-1.2}$ & $1.0^{+1.3}_{-0.2}$\\
\tableline
\end{tabular}
\vspace*{\baselineskip}~\\ \end{center} \tablecomments{Results of
  fitting simple models to the EPIC spectra of four bright ULXs
  in the Antennae galaxies.  The XSPEC model ``phabs''was used to measure the
  equivalent neutral hydrogen column density along the line of
  sight.\\   $^{a}$ The measured flux.\\
$^{b}$ The absorption-corrected or ``unabsorbed'' flux.\\
  $^{c}$ The source luminosity in the 0.3--10.0~keV band for a
  distance of 19~Mpc.}
\vspace{-1.0\baselineskip}
\end{footnotesize}
\end{table}

\clearpage

\centerline{~\psfig{file=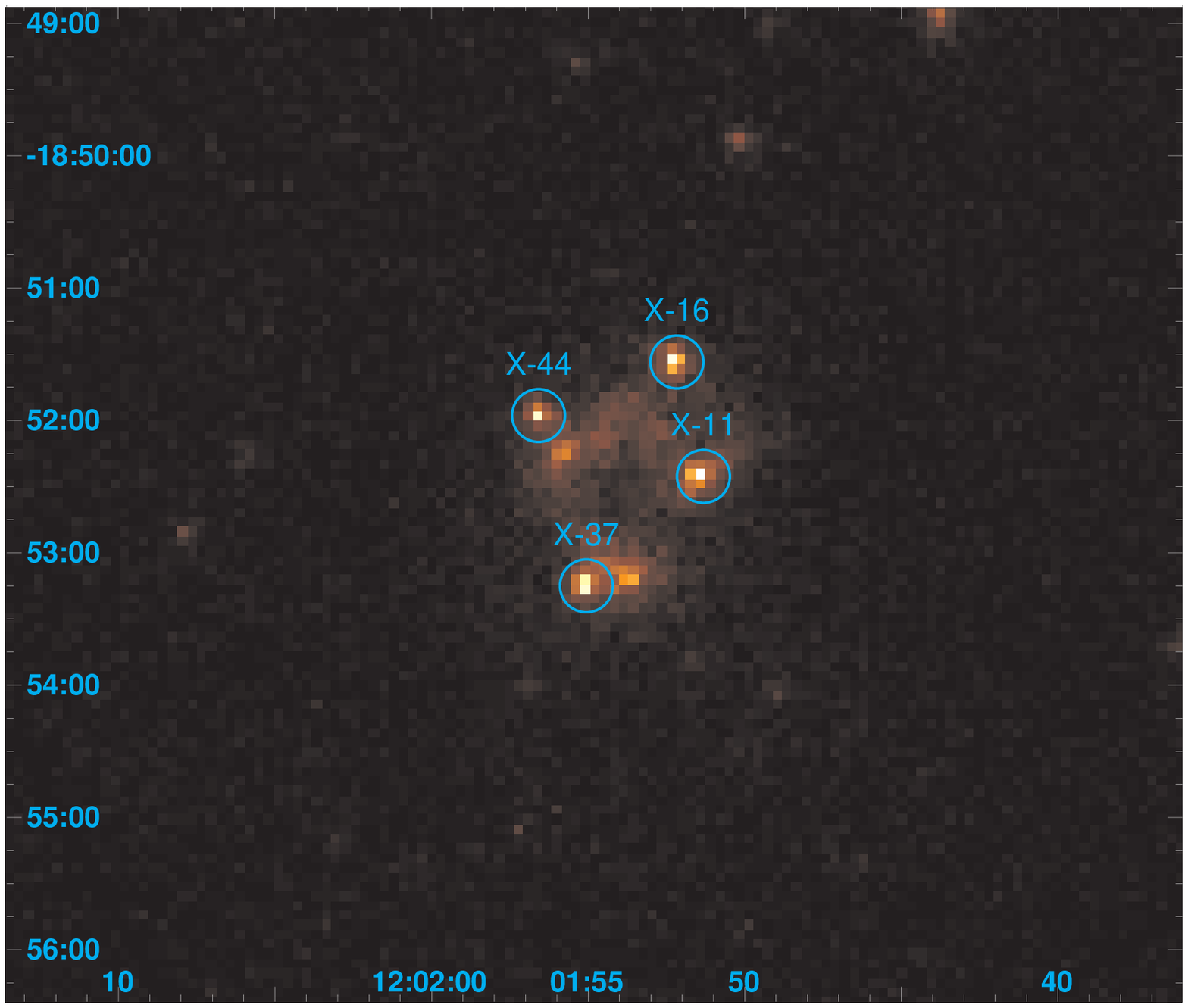,width=3.5in}~}
\figcaption[h]{\small The EPIC/MOS1 0.3-10.0~keV image of the central
  regions of the Antennae galaxies, shown on a linear scale.  The
  source spectra considered in this work were extracted using the
  circular regions shown, which are centered on previously-measured
  {\it Chandra} source positions.}
\medskip

\clearpage

\centerline{~\psfig{file=f2.ps,width=4.5in,angle=-90}~}
\figcaption[h]{\small The pn (black), MOS-1 (red), and MOS-2 (blue)
spectra of X-11 fit with a simple power-law model (see Table 1).}
\medskip

\centerline{~\psfig{file=f3.ps,width=4.5in,angle=-90}~}
\figcaption[h]{\small The pn (black), MOS-1 (red), and MOS-2 (blue)
spectra of X-16 fit with a simple power-law model (see Table 1). }

\clearpage

\centerline{~\psfig{file=f4.ps,width=4.5in,angle=-90}~}
\figcaption[h]{\small The pn (black), MOS-1 (red), and MOS-2 (blue)
spectra of X-37 fit with a simple power-law model (see Table 1).}
\medskip

\centerline{~\psfig{file=f5.ps,width=4.5in,angle=-90}~}
\figcaption[h]{\small The pn (black), MOS-1 (red), and MOS-2 (blue)
spectra of X-44 fit with a simple power-law model (see Table 1).}

\clearpage

\centerline{~\psfig{file=f6.ps,width=4.5in,angle=-90}~}
\figcaption[h]{\small The best-fit models for each ULX source (see
  Table 1).  In black: the power-law model for X-11.  In red: the
  power-law model for X-16.  In green, the MCD plus power-law
  model for X-37 (with disk and power-law components also shown
  in green).  In red, the bremsstrahlung model fit to X-44.
  Clearly, the best-fit models for the ULX spectra we consider are
  significantly different; this may indicate different source types.}
\medskip

\end{document}